\documentclass[preprint,pre,floats,showkeys,aps,amsmath,amssymb,12pt]{revtex4}
\usepackage{graphicx}
\usepackage{natbib}
\usepackage{amsthm}
\usepackage{enumerate}% http://ctan.org/pkg/enumerate
\usepackage[left = 1in, top=1.2in,right=1in,bottom=1.2in,a4paper]{geometry}
\usepackage{amsmath}
\usepackage{amssymb}
\usepackage{hyperref}
\newcommand{\ket}[1]{|#1\rangle}
\newcommand{\bra}[1]{\langle #1|}

\begin{document}

%\markboth{Mishkat Al Alvi, Md. Abdul Matin (deceased), Moinul Hossain Rahat, Avik Roy, and Mahbub Majumdar}
%{Comments on Information Erasure in Black Hole}
%
%%%%%%%%%%%%%%%%%%%%%% Publisher's Area please ignore %%%%%%%%%%%%%%
%\catchline{}{}{}{}{}
%%%%%%%%%%%%%%%%%%%%%%%%%%%%%%%%%%%%%%%%%%%%%%%%%%%%%%%%%%%%%%%%%%%%

\title{Comments on Information Erasure in Black Hole \\
%USING \TeX\ OR \LaTeX\footnote{For the title, try not to use more than
%three lines. Typeset the title in 10 pt Times Roman, uppercase and
%boldface.}
}

\author{Mishkat Al Alvi \footnote{Email: mishkat.alvi@gmail.com}, Md. Abdul Matin (deceased), Moinul Hossain Rahat \footnote{Email: rahat.moin90@gmail.com}, Avik Roy \footnote{Email: avik\_3.1416@yahoo.co.in}}
\affiliation{Department of Electrical \& Electronic Engineering,\\ Bangladesh University of Engineering \& Technology\\
Dhaka - 1000,
Bangladesh}

%
%\author{}
%\affiliation{Department of Electrical \& Electronic Engineering,\\ Bangladesh University of Engineering \& Technology\\
%Dhaka - 1000,
%Bangladesh \\
%}
%
%\author{}
%\affiliation{Department of Electrical \& Electronic Engineering,\\ Bangladesh University of Engineering \& Technology\\
%Dhaka - 1000,
%Bangladesh \\ }
%\&
\author{Mahbub Majumdar \footnote{Email: majumdar@bracu.ac.bd}$^{,}$\footnote{Email: mahbub@thematrics.org}}
\affiliation{BRAC University, 66 Mohakhali, Dhaka - 1212,
Bangladesh }
\affiliation{Theoretical Physics \& Mathematics Groups \\ The Mathematics \& Arts, Teaching \& Research Institute for Culture \& Science (MATRICS), \\ Abdul Monem Business District, 111 Bir Uttam C. R. Datta Road, Dhaka - 1205, Bangladesh \vspace{0.4in}}

%\maketitle

%\pub{Received (Day Month Year)}{Revised (Day Month Year)}

\begin{abstract}
We analyze the Kim, Lee \& Lee model of information erasure by black holes and find contradictions with standard physical laws. We demonstrate that the erasure model leads to arbitrarily fast information erasure; the proposed physical interpretation of information freezing at the event horizon as observed by an asymptotic observer is problematic; and information erasure, whatever the process may be, near the black hole horizon leads to contradictions with quantum mechanics if Landauer's principle is assumed. The later part of the work demonstrates the significance of the ``erasure entropy."  We show that the erasure entropy is the mutual information between two subsystems.
\end{abstract}

%\ccode{PACS Nos.: include PACS Nos.}
\maketitle

\section{Introduction}
Landauer's principle [\cite{landauer}] states that the erasure of 1 bit of information increases the classical entropy by
\begin{equation} \label{eqn:landauer}
\Delta S \ge k \log{2}
\end{equation}
where $k$ is Boltzman's constant. Landauer's Principle has been important in understanding the thermodynamics of quantum computation [\cite{lan2},\cite{lloyd},\cite{Janzing},\cite{bennet}].

Extensions to quantum systems has inspired many explorations. For example, the thermodynamic costs of processing classical information encoded in quantum systems was discussed by [\cite{Plenio2}]. In relating Landauer's principle to von Neumann entropy, Plenio reformulated the Holevo bound [\cite{holevo}, \cite{plenio}].

Explicit relations between thermodynamics and quantum information via Landauer's principle has spurred applications to black hole thermodynamics. It has been used to derive a generalized second law for black hole thermodynamics [\cite{song}]. Recently, it has been used to try to resolve the black hole information paradox [\cite{eraser},\cite{herrera}] by relating the black hole entropy to an erasure entropy.  In this paper, we produce counterarguments to the erasure process in [\cite{eraser}] and find contradictions with standard results. We then argue that it is not possible for any erasure process to erase information near the black hole horizon if Landauer's principle is assumed to hold. We also clarify the notion of quantum information erasure as mentioned in the literature.

The paper is organized as follows. Section \ref{sec:review} reviews the information erasure model advocated in [\cite{eraser}]. Section \ref{sec:incon} details the inconsistencies in the erasure model and demonstrates contradictions between the erasure model and standard results. In section \ref{sec:impossibility}, we explain why no erasure process, irrespective of the details, can be consistent with Landauer's principle. Then in section \ref{sec:general}, the general meaning of quantum information erasure is explained. It is showed that the so-called erasure entropy is the mutual information of two systems. Finally in section \ref{sec:conclusion}, the central ideas of the paper are summarized.

\section{Review of information erasure proposed by Kim, Lee \& Lee}\label{sec:review}

Standard lore states that a faraway observer never sees an infalling particle cross a black hole apparent/event horizon. From this viewpoint, it takes an infinite amount of time for an infalling particle to cross the horizon as the waves radiated by the particle experience gravitational redshift. This is called `information freezing' at the horizon. Freezing is considered to be a coordinate artifact.

Kim, Lee and Lee proposed in [\cite{eraser}] that information freezing might correspond to a physical process. They suggested that the information of the infalling particle is first frozen at the horizon and then erased by interaction with future infalling particles. The energy required for information erasure comes from the energy of infalling matter.  This energy and entropy is absorbed inside the black hole.

The authors assume that the black hole event horizon is shielded by a `system' of finite energy, so that this hypothetical system is invisible to asymptotic observers because of severe redshifting. The process of infall is divided into three parts.

\begin{enumerate}[i.]
\item The energy of the particle erases the information in the system. This erasure of information is associated with a change in entropy as dictated in [\ref{eqn:landauer}].
\item The black hole horizon increases and engulfs the `system.' The energy of the particle and entropy of erasure is consumed by the black hole.
\item The information of the infalling particle is frozen at the horizon and becomes a part of the new system.
\end{enumerate}

\section{Inconsistencies in the Kim, Lee \& Lee model}\label{sec:incon}
Kim, Lee \& Lee postulate an eraser system at the black hole horizon.  They do not offer many details on how the system physically stores or erases information. Their hypothetical system is however problematic at a conceptual level, as is shown by the following general arguments.

Let a small mass particle fall towards a black hole. Assume that the infalling particle increases the mass of a Schwarzschild black hole from $M$ to $M + \Delta M$.  It therefore increases the radius $R$ of the event horizon by $\Delta R$.  Assume that the black hole's temperature $T = 1/(8\pi k M)$, does not change significantly.  This is a good assumption for large mass black holes. From the first law of black hole thermodynamics, the black hole's mass and radius changes are

\begin{eqnarray}
 \Delta M = \frac{k\log 2}{8\pi M}; & {\;\;\;\;\;\;}& \Delta R \le \frac{k\log 2}{2 \pi R}
 \label{eqn:3}
\end{eqnarray}

\noindent
where the inequality on the right follows from the supposition that $\Delta R \le 2 \Delta M$.

For simplicity and without loss of generality, assume that the shell of width $\Delta R$ comprises the `system' as defined by Landauer and Kim, Lee \& Lee.  Assume it encodes 1 bit of information.

Since the infalling particle has negligible mass, we can assume it doesn't perturb the geometry significantly. Let the proper time along the world line of the particle be $\tau$. Assume the particle started from infinity with zero initial velocity.  Therefore the particle's proper time $\tau$ asymptotically becomes coordinate time $t$. Suppose the particle is at $r_0 = R + \Delta R$ at proper time $\tau_0$, and at $r = R$ at proper time $\tau$. Standard results show that [\cite{dinverno}]

\begin{equation}\label{eqn:din1}
  \Delta \tau = \tau - \tau_0 = \frac{2}{3\sqrt{2M}} (r_0^{3/2} - r^{3/2})
\leq \frac{k}{2\pi R}\log 2
\end{equation}

This gives the interaction time between the system and the infalling photon. The time $\Delta \tau$ can be made arbitrarily small as $M$ increases.

In the framework of the Kim, Lee \& Lee model, and given the freefall assumption, $\Delta \tau$ represents the time required for information erasure from the system.  From (\ref{eqn:din1}) we see that this rate of (classical) information erasure can be made arbitrarily large.  This is problematic.

For the sake of completeness of our argument, consider quantum effects.  Assume the system is entangled with degrees of freedom inside and outside the event horizon. We do not know the detailed information erasure mechanism, but we know that the evolution of the entire system is unitary. Unitary evolution preserves entropy implying that information cannot be destroyed at the quantum level [\cite{pati2}]. However, information can be  `erased' from a part of a system by transferring it to another component of the system, which is what Kim Lee \& Lee are essentially describing.

Their  model can be regarded as representative of models of information processors near the horizon that suggest ways around the information paradox or other ways to interpret the information paradox. A possible way to physically realize/describe the Kim, Lee \& Lee model is as follows.

Consider a simple qubit example that implements the desired information transfer. Denote the `system' as $A$, the environment it is entangled with as $B$ and the infalling particle as $C$. Additionally, assume that both $A$ and $C$ are simple single qubit spaces and encode a maximum of $1$ bit of information. We desire a unitary transformation $\hat{U}$ of the form

\begin{equation}
\hat{U}\{\hat{\rho}_{AB}\otimes\hat{\rho}_C\}\hat{U}^\dagger = \hat{\rho}_{ABC}
\end{equation}

\noindent so that the reduced density matrix of the system after the transformation is the same as the density matrix of  the infalling particle before the transformation,

\begin{equation}
\mathrm{tr}_{B,C}\left(\hat{\rho}_{ABC}\right) = \hat{\rho}_C
\end{equation}

This task can be accomplished for example, by applying a quantum bit swap gate %add a reference?!?
to the states of $A$ and $C$ while leaving $B$ undisturbed. Only the `system' interacts with the infalling particle -- a natural requirement.

%Any such protocol is inherently invalidated by the equivalence principle. An infalling observer should find the horizon to be innocuous, in the Hartle-Hawking vacuum state. The event horizon is a regular place for an infalling observer so nothing out of ordinary should take place at the horizon.

%We are trying to address the same issue from a quantum mechanical perspective. In particular, we will show why a system with finite energy and constrained physical dimension is at odds with general quantum mechanical arguments.

Let us assume that the `system' proposed by Kim, Lee \& Lee uses some information processing protocol near the horizon. Assuming unitarity, the time evolution will be generated by a unitary operator $\hat{U} \ne \hat{I}$ that is nontrivial

\begin{equation}
\hat{U} = \exp\left(-i\hat{H}\Delta\tau\right),
\end{equation}
where $\hat{H}$ is the Hamiltonian of the joint system.

Now $\Delta\tau$ which is given by (\ref{eqn:din1}), can be made arbitrarily small for large black holes. If we assume the energy of the `system,' which is a shell or size $\Delta R$ near the horizon, is finite (severe redshifting makes this energy invisible to asymptotic observers), then we can achieve $\hat{U} \sim \hat{I}$. Thus the system remains almost unchanged.

%The increased radius of the black hole due to absorption of energy will cause the information in the system to enter the black hole.

Another way to compensate for infinitesimally small $\Delta\tau$ is to allow the energy eigenvalues of the system to become arbitrarily large.  But, then this process of energy erasure at the horizon would no longer be invisible to asymptotic observers. Hence, the claim that information erasure is a `physical realization' of `information freezing,' which is invisible to asymptotic observers, would become troublesome at best in the context of the Kim, Lee \& Lee model. No unitary evolution of an information containing `system' engaged in the three step erasure process can erase (or even process) quantum information without being noticed by asymptotic observers.

We have thus showed two significant loopholes in the erasure process proposed in [\cite{eraser}].

\begin{enumerate}[i.]
\item Classical information can be erased arbitrarily quickly from the `system'.
\item There is no unitary evolution that erases quantum information under the assumption of finite energy of the `system.' With the finite energy condition relaxed, this erasure process is no longer a `physical realization' of `information freezing' at the horizon.
\end{enumerate}

\section{Impossibility of information erasure processes near a black hole horizon}\label{sec:impossibility}

In this section we consider more general arguments about information erasure near a black hole event horizon. Some authors have speculated   that the information paradox may be resolved by bleaching information near the horizon [\cite{eraser},\cite{herrera}].

Let us consider some general process of information erasure near the black hole horizon that respects the uncertainty relations and Landauer's principle. Erasing one bit of information has an entropic cost of at least $k\log 2$.  For large black holes which experience negligible temperature change, the energy change associated with this process is given by

\begin{equation}
\Delta E \ge k T_{BH}\log 2
\end{equation}

\noindent
where $T_{BH}$ is the black hole's temperature and satisfies $T_{BH} = 1/8\pi kM$.  Employing the energy-time uncertainty relation, i.e. $\Delta E \ \cdot \Delta \tau \sim 1$ we find

\begin{equation} \label{eqn:10}
\Delta \tau \sim \frac{8\pi M}{\log 2}.
\end{equation}

The time required for information erasure depends on the computational efficiency of a black hole. Equation (\ref{eqn:10}) implies that larger black holes require more time for a single bit operation like erasure.

The time for processing one bit of information should be constrained by the minimal resolvable timescale by a black hole.  This is proportional to the black hole mass as derived in  [\cite{barrow},\cite{yng}]. The derivation in [\cite{barrow}] involves using clock inequalities as used by Wigner [\cite{wigner}].  These are basically modified representations of the energy-time uncertainty principle, and are thus similar to the argument used in deriving (\ref{eqn:10}).

The result in (\ref{eqn:10}) implies that $f_{BH}$, the number of computational operations per bit per unit time for a black hole is  $f_{\mathrm{BH}} \sim \frac{1}{\Delta \tau} \sim \frac{1}{M}$. Since a black hole's entropy is proportional to its event horizon area, the number of bits it can encode -- the size of its memory space $\mathcal{S}_{BH}$ -- is proportional to its area $\mathcal{A}_{BH}$, which is proportional to $M^2$.  A black hole is often thought of as a serial computer becomes it computes so fast. If we enable parallelization, so that processing can occur on each bit in the memory space, then $N$, the number of operations per unit time for a black hole computer is $N = f_{BH} \times \mathcal{S}_{BH} \sim M$. This is in concordance with Lloyd [\cite{Lloyd1}]:  the {\em total} number of bit operations per unit time by an information processing system of finite energy is proportional to the energy of the system.
%However, The proportionality between the maximum number of operations possible and black hole mass appears from the memory space of a large black hole which is proportional to the area of the event horizon and hence $M^2$. Time required for a single bit operation, as identified in (\ref{eqn:10}) and also investigated in Refs.~\cite{barrow} and \cite{yng}, is proportional to the black hole mass. The derivation in Ref.~\cite{barrow} involved using the clock inequalities by Wigner \cite{wigner}, which are basically modified representations of the energy-time uncertainty principle, the same argument used in deriving (\ref{eqn:10}).$

The result in (\ref{eqn:10}) is also in concordance with results obtained by Pfeifer [\cite{Pfeifer}] and Margolus et al. [\cite{margolus}] with regard to the minimum timescale of evolution of a quantum system.  In [\cite{Pfeifer}] the minimum time for a quantum state to evolve to an orthogonal state is bounded by

\begin{equation} \label{eqn:star1}
\Delta \tau \geq \frac{h}{4 \Delta E}.
\end{equation}

\noindent
In [\cite{margolus}] the same quantity was bounded by

\begin{equation} \label{eqn:star2}
\Delta \tau \geq \frac{h}{4(E-E_0)}
\end{equation}

\noindent
where $E$ is the average energy of the system and $E_0$ is the ground state energy.

These different relations apply best to different regimes. For systems with finite average energy and large standard deviations, (\ref{eqn:star2}) serves as a better bound. For systems with large average energy but small standard deviations like black hole systems, (\ref{eqn:star1}) is the better estimate.

 A simple example can illustrate this. Consider the state $\ket{w} = a\ket{0} + b\ket{E}$ where $|b| \gg |a|$ and $\ket{E}$ and $\ket{0}$ are energy eigenstates. Margolus and Levitin's methods show that the system will never evolve to an orthogonal system; the bound (\ref{eqn:star2}) therefore does not provide meaningful information. The bound in (\ref{eqn:star1}) is meaningful in the limiting case, $|a| \rightarrow 0$ and $|b| \rightarrow 1$ since the timescale becomes arbitrarily large which corresponds to physical reality.  The timescale (\ref{eqn:star1}) is also consistent with (\ref{eqn:10}) where Landauer's principle is invoked to estimate the energy uncertainty associated with a single qubit operation.

%The bounds (\ref{eqn:star1}) and (\ref{eqn:star2}) represent general lower bounds on time required for a quantum system to evolve to an orthogonal state.

 The actual timescale to evolve to an orthogonal state is subject to physical constraints, not just (\ref{eqn:star1}) or (\ref{eqn:star2}). For black holes, the minimum resolvable time limits the speed of possible operations. The proportionality between $\Delta \tau$, interpreted as the interaction time associated with the uncertainty of energy, and black hole mass is evident from (\ref{eqn:10}). This is in clear contradiction with the formula for $\Delta \tau$ in (\ref{eqn:din1}).

Let us us see how compatible the estimate for $\Delta \tau$ in (\ref{eqn:10}) is with the assumption that particles free falling through the $\Delta R$ shell around horizon cause information erasure.

Define the shell $R\le r \le R+ \Delta R$ as the region of information erasure,  where $\Delta R \ll R$. Some mechanism in this region causes the information to be erased. The process of erasure increases the black hole's entropy via Landauer's principle.  If the width $\Delta R$ were large the change of entropy could be observed in the surrounding environment. The size of $\Delta R$ should be determined by the small perturbations to the Schwarzschild metric. A similar idea has been advocated in [\cite{preskill}] where the so-called `stretched horizon' acts as a `repository that stores quantum information absorbed by the black hole' and has a width of the order of planck length.

Now, if the information bearing particle follows a timelike geodesic in approaching the black hole, the proper time it requires is given by (\ref{eqn:din1}). With the approximation $\Delta R \ll R$, this may be reduced to

\begin{equation}
\Delta \tau \sim \Delta R \ll R
\end{equation}

This contradicts (\ref{eqn:10}) which stipulates $\Delta \tau \sim R$ and which is based on Landauer's Principle.  This contradiction seems generic to any process that advocates information erasure near the black hole.  It therefore prompts  a more general result -- ``Information erasure processes that respect Landauer's principle cannot erase the information of an infalling particle near the black hole event horizon."

%In fact, this result is equivalent to the equivalence principle in Einstein's theory of general relativity to some extent, in concord with the fact that the horizon is an innocuous place for the infalling observer and nothing out of ordinary should take place in the local reference frame of the infalling observer.

In general, several authors [\cite{preskill}, \cite{page}] have argued that quantum information processing in a black hole takes a `considerable' amount of time. For example, [\cite{preskill}] advocates that `thermalization' of infalling information takes a time of order $R \log R$. These arguments support our claim that information processing like erasure cannot occur arbitrarily quickly (\ref{eqn:din1}).

\section{General meaning of information erasure in quantum context}\label{sec:general}

%Expansion of Landauer's principle in the context of quantum mechanics has been elaborated in literature by various authors. \cite{lubkin, plenio, song} While erasure is absolutely irrelevant in context of quantum information, \cite{pati2} since quantum systems evolve unitarily and hence preserves information, such explorations in literature needs specific clarification.

After information is erased near the horizon, where does it go?  Quantum mechanically, information is not erased. It is instead transferred to another part of the system.  In this section we provide an interpretation of information erased through a Landauer-like process to  understand where the erased information appears.  We  present arguments in [\cite{song}] which extend  arguments in [\cite{lubkin},\cite{plenio}].

Consider a black hole surrounded by a spherical cavity of comparatively large dimensions. Radiation fills the cavity and is in thermal equilibrium with the black hole. We treat this radiation filled cavity as a thermal reservoir.  Suppose a quantum system is in the state

\begin{equation}
\ket{\psi_S} = \sum_i \sqrt{\lambda_i}\ket{a_i}.
\end{equation}

Also, suppose that the apparatus $M$  is initially in a pure state.  During the process of measurement this apparatus entangles itself with the system and the joint state of the apparatus and the system is given by

\begin{equation}
\ket{\psi_{SM}} = \sum_i \sqrt{\lambda_i}\ket{a_i}_S\ket{m_i}_M
\end{equation}

\noindent
where \{$| m_i \rangle $\} is  basis of states for the apparatus.

The density matrix describing the state of the apparatus is given by

\begin{equation}\label{rhom}
\hat{\rho}_M = \sum_i\lambda_i\ket{m_i}\bra{m_i}.
\end{equation}

\noindent
The apparatus is thus found in a state $\ket{m_i}$ with probability $\lambda_i$. After a measurement takes place, the apparatus will be in a pure state. The general way to erase this information is to bring the apparatus in thermal contact with a heat reservoir of temperature $T$ which projects the apparatus'  state to a thermal state irrespective, thus erasing the information [\cite{song}].  In [\cite{pati}] such processes that project every state to a fixed state (of some other subspace in general) have been termed `information hiding processes.'

If $\hat{H}$ is the hamiltonian of the apparatus, the density operator of the apparatus in thermal equilibrium with the reservoir is given by

\begin{equation}
\hat{\omega}_M = \frac{e^ {-\hat{H}/k T}}{Z}
\end{equation}

\noindent
where $Z$ is the thermal partition function. After erasing the information, the apparatus gains an entropy

\begin{equation}\label{delapp}
\Delta S_{app} = -\mathrm{tr}\left(\hat{\omega}_M\log{\hat{\omega}_M}\right).
\end{equation}

Assuming that the heat reservoir's temperature is unchanged in the process, the change in its entropy, in dimensionless units is given by the classical thermodynamic relation

\begin{eqnarray}\label{delsr}
    \Delta S_R = \frac{\Delta Q}{k T}
   & = & \frac{\mathrm {tr}\{(\hat{\rho}_M-\hat{\omega}_M)\hat{H}\}}{k T}
    = -\mathrm{tr}\{(\hat{\rho}_M-\hat{\omega}_M)\log\hat{\omega}_M\}
\end{eqnarray}

\noindent
where in the final step we used $\mathrm{tr}\{(\hat{\rho}_M-\hat{\omega}_M) \log Z\} = 0$, which follows from probability conservation.

The total entropy change from erasure $\Delta S_{era}$, is given by the sum of the entropy changes of the reservoir (\ref{delsr}) and the measuring apparatus (\ref{delapp}),

\begin{equation}\label{sera}
\Delta S_{era} = \Delta S_{app} + \Delta S_{R} = -\mathrm{tr}\{\hat{\rho}_M\log\hat{\omega}_M\} \ge 0
\end{equation}

\noindent
which is positive via Klein's inequality: $\mathrm{tr}\{\hat{\rho}(\log{\hat{\rho}}-\log{\hat{\omega}})\}\ge 0$.

Quantum mechanically the erased information cannot simply disappear. It is therefore not surprising that this erasure entropy $\Delta S_{era}$ ends up in the correlations between the apparatus $M$ and the heat bath reservoir $R$ as we show below.

Suppose $\hat{\rho}_R$ is the density matrix of the heat reservoir and the apparatus is found in a pure state with density matrix $\hat{\rho}_{M} = \ket{m_i}\bra{m_i}$  before thermalization of the apparatus. The thermalization process is described by some unitary matrix over the joint system that entangles the apparatus and reservoir,

\begin{equation}
\hat{\rho}_R \otimes \hat{\rho}_{M} \to \hat{\rho}_{RM}.
\end{equation}

The final state of the apparatus is obtained by tracing over the reservoir system and the final state of the reservoir is similarly given,

\begin{eqnarray}
\hat{\omega}_M = \mathrm{tr}_R \{\hat{\rho}_{RM}\}; & \;\;\;&
\hat{\omega}_R = \mathrm{tr}_M \{\hat{\rho}_{RM}\}.
\end{eqnarray}

Unitarity ensures that no information is lost. This implies $S\left(\hat{\rho}_R\otimes \hat{\rho}_{M}\right) = S\left(\hat{\rho}_{RM}\right)$. The entangled reservoir-apparatus system will encode some of the information initially in the separate $R$ and $M$ systems in the correlations between $R$ and $M$.  This is given by the mutual information $S(R:M)$ between the heat reservoir and apparatus. It is zero if no correlations between $R$ and $M$ exist.
\begin{align}
S(R:M) &= S(\hat{\omega}_R) + S(\hat{\omega}_M) - S(\hat{\rho}_{RM}) \nonumber
\\ &= S(\hat{\omega}_R) + S(\hat{\omega}_M) - \left(S(\hat{\rho}_R)+S(\hat{\rho}_M)\right)     \nonumber
\\ & = [S(\hat{\omega}_R)-S(\hat{\rho}_R)] + [S(\hat{\omega}_M) - S(\hat{\rho}_M)]  \nonumber
\\ & = \Delta S_{R} + \Delta S_{app}    \nonumber
\\ & = \Delta S_{era}
\end{align}

The erasure entropy $\Delta S_{era}$ is therefore shifted into the correlations of the subsystems after information erasure.

\section{Conclusion}\label{sec:conclusion}
This work examines the feasibility of the Kim, Lee \& Lee proposal [\cite{eraser}] which suggests that the black hole information paradox may be resolved by the erasure of information near the horizon. This is a very interesting proposal, but we find several weaknesses in their prescription.

Assuming the horizon is smooth and particles free-fall through it, in the Kim, Lee \& Lee model, information erasure near the horizon can happen arbitrarily quickly.  Also, we show that finite energy processes cannot cause a significant change in the quantum state of an infalling particle, thus precluding meaningful information erasure at the horizon. In the next section, we showed that no erasure process conforming to Landauer's principle can erase quantum information in the vicinity of the black hole horizon without violating an uncertainty principle.  We then showed that the erased entropy appears in the mutual information between two subsystems. This illustrates a major difference between classical and quantum information erasure processes. Erased classical information is irretrievable, but quantum mechanically it still lives on in subsystem correlations.

\bibliography{erasure}
\bibliographystyle{unsrt}
%\begin{thebibliography}{0}
%\bibitem{Marnelius} R. Marnelius, {\it Acta Phys. Pol. B} {\bf 13},
%669 (1982).
%
%\bibitem{Bjorken} J. D. Bjorken, in {\it Lecture Notes on Current-Induced
%Reactions}, eds.~J. Komer {\it et al.} (Springer, 1975).
%
%\bibitem{Bohr} A. Bohr and B. R. Mottelson, {\it Nuclear Structure}
%(Benjamin, 1969), Vol.~1, pp.~100--102.
%
%\bibitem{Webb} R. C. Webb, Ph.D. thesis, Princeton University, 1972.
%
%\bibitem{Toimela} T. Toimela, Helsinki Research Institute for
%Theoretical Physics, Report No. HU-TFT-82-37, 1982 (unpublished).
%\end{thebibliography}
\end{document}